\documentclass[aps,prl,twocolumn,groupedaddress,showpacs,showkeys]{revtex4-1}

\usepackage{natbib}
\usepackage{floatflt}
\usepackage[dvips]{graphicx}
\usepackage{dcolumn}
\usepackage{bm}
\usepackage[all]{xy}
\usepackage{amssymb}
\usepackage{amsmath}

\begin{document}

\preprint{AR-4}
\title{Noise reduction by coupling of stochastic processes and canalization in biology} 

\author{Alexandre F. Ramos} \affiliation{Escola de Artes, Ci\^encias e
  Humanidades, Universidade de S\~ao Paulo, Av. Arlindo B\'ettio, 1000
  CEP 03828-000, S\~ao Paulo, SP, Brazil} \email{alex.ramos@usp.br}
\author{Jos{\'e} Eduardo M. Hornos} \affiliation{Instituto de
  F\'{\i}sica de S\~ao Carlos, Universidade de S\~ao Paulo, Caixa
  Postal 369, BR-13560-970 S\~ao Carlos, S.P., Brazil} \author{John
  Reinitz} \affiliation{Department of Statistics, Department of
  Ecology and Evolution, Department of Molecular Genetics and Cell
  Biology, University of Chicago 5734 S. University Avenue Eckhart 134
  Chicago, IL 60637, USA}

\maketitle

\date{today}

Randomness is an unavoidable feature of the intracellular environment
due to chemical reactants being present in low copy number. That
phenomenon, predicted by Delbr\"uck long ago \cite{delbruck40}, has
been detected in both prokaryotic \cite{elowitz02,cai06} and
eukaryotic \cite{blake03} cells after the development of the
fluorescence techniques.  On the other hand, developing organisms,
{\em e.g.} {\em D. melanogaster}, exhibit strikingly precise
spatio-temporal patterns of protein/mRNA concentrations
\cite{gregor07b,manu09a,manu09b,boettiger09}. Those two characteristics of
living organisms are in apparent contradiction: the precise patterns
of protein concentrations are the result of multiple mutually
interacting random chemical reactions. 
The main question is to establish 
biochemical mechanisms for coupling random reactions so that
canalization, or fluctuations reduction instead of amplification,
takes place. Here we explore a model for coupling two stochastic
processes where the noise of the combined process can be smaller than
that of the isolated ones. Such a canalization occurs if, and only if,
there is negative covariance between the random variables of the
model. Our results are obtained in the framework of a master equation
for a negatively self-regulated -- or externally regulated -- binary
gene and show that the precise control due to negative self regulation
\cite{becskei00} is because it may generate negative covariance.
Our results suggest that negative covariance, in the coupling of
random chemical reactions, is a theoretical mechanism underlying the
precision of developmental processes.

We approach the stochastic model for a binary gene operating under
negative self-regulation or external regulation \cite{sasai03,
  peccoud95}. Analytic solutions for the steady state
\cite{Hornos05,innocentini07} and the dynamic \cite{ramos11, biswas09}
regimes, as well as the symmetries \cite{ramos07, ramos10} underlying
solubility have already been presented. This system has the
transcription and translation treated as combined processes. The state
of the system is defined by two stochastic variables, the gene state
(activate or repressed) and the protein number in the cytoplasm. The
gene state is defined effectively in terms of its promoter site as
on/off under the action of an external agent, {\em e.g.} a protein
codified by another gene, or by self-interaction.

We consider a stochastic formulation for the dynamics of the
probability of finding the gene in an active (or repressed) state,
indicated by $\alpha_n$ (or $\beta_n$) when $n$ proteins are found in
the cytoplasm. The protein synthesis rates are given by $k$ or $\chi
k$ ($0\le\chi<1$) when the gene is, respectively, turned ``on'' or
``off''. The degradation rate of the proteins is given by $\rho$. The
``off-on'' switching rate is denoted by $f$ while $h_1$ and $h_2$
indicate the opposite transition rates. The master equation is:
\begin{eqnarray}
\dfrac{d\alpha_n}{dt} &=& k (\alpha_{n-1}-\alpha_n) +
\rho[(n+1)\alpha_{n+1}-n\alpha_n] \nonumber \\ &-& ({h}_1 n+ {h}_2)
\alpha_n + {f} \beta_n, \label{equa1}
\end{eqnarray}
\begin{eqnarray}
\dfrac{d\beta_n}{dt}  &=& \chi k (\beta_{n-1}-\beta_n) 
+ \rho[(n+1)\beta_{n+1}-n\beta_n] \nonumber \\
 &+& ({h}_1 n+ {h}_2) \alpha_n   - {f} \beta_n. \label{equa2}
\end{eqnarray}
The existence of an ``on-off'' transition dependent on $n$ indicates a
negative self-regulating gene ($h_2=0$). For an externally regulated
gene, one assumes $h_1$ to be zero. The proportionality to $n$ is
effective and has no relationship with the actual biochemical
mechanisms of protein binding/unbinding to DNA regulatory regions,
that might involve a plethora of chemical reactions. 

Eqs. (\ref{equa1}) and (\ref{equa2}) might be considered as the
coupling of two different Poissonian processes, each of them related
to one of the gene states. The processes are coupled in terms of a
second stochastic variable, the gene state. To each gene state, we
associate the quantities $N_1=k/\rho$ and $N_2=\chi k/\rho$, that are
the stationary averages of the isolated Poisson processes.  Therefore,
the biologically measurable quantities are the protein number in the
cytoplasm $n$ and the protein synthesis rates $\{N_1, N_2\}$ and proceed to evaluating the noise on $n$.

For that purpose, we start defining the moments of the random
variables $(n,N)$ in terms of $\alpha_n$ and $\beta_n$ as:
\begin{equation}\label{moments}
\langle n^p N^q \rangle = \displaystyle \sum_{n=0}^{+\infty} (\alpha_n
N_1^q + \beta_n N_2^q) n^p.
\end{equation}
The marginal probabilities of finding $n$ proteins inside the cell are
given by $\phi_n = \alpha_n + \beta_n$, and the protein synthesis rate
has a probability $p_1 = \sum_{n=0}^\infty \alpha_n$ (or $p_2 =
\sum_{n=0}^\infty \beta_n$) to be $N_1$ (or $N_2$).

As we have two random variables, it is convenient to use the
covariance between $n$ and $N$ -- indicated by $\xi_{i,n}$ -- for the
analysis of the variance on the number of gene products, namely
\begin{equation} \label{covar}
\xi_{n,i} = \langle n N \rangle - \langle n \rangle \langle N \rangle.
\end{equation}

The noise on the protein number of the composed system of the
Eqs. (\ref{equa1}) and (\ref{equa2}) is computed in terms of the Fano
factor, that is defined as the ratio between the variance and the mean
of $n$,
\begin{equation}
\mathcal{F} = \frac{\langle n^2 \rangle - \langle n \rangle^2}{\langle
  n \rangle}.
\end{equation}
As a Poisson (or Fano) distribution has a Fano factor equals to unity,
the Fano factor is used to determine how different from the Poissonian
a probability distribution is. When $\mathcal{F} > 1$ the distribution
is spreader and named super-Poisson (or super-Fano) while it is named
sub-Poisson (or sub-Fano) when $\mathcal{F} < 1$. As it is shown at
the supplementary material, at the stationary limit, the Fano factor
can be reduced to 
\begin{equation} \label{fano}
\mathcal{F} = 1 + \frac{\xi_{n,i}}{\langle n \rangle},
\end{equation}
which value depends on the signal of the covariance between the two
stochastic variables of the model. It is worth to mention that this
relation for the Fano factor holds for a gene operating in two, three
and so on, states of synthesis.

Eq. (\ref{fano}) shows the possibility of occurrence of sub, super or
Poissonian distributions when we deal with the probability
distributions generated by the Eqs. (\ref{equa1}) and (\ref{equa2}),
depending on the value of $\xi_{n,i}$. As we shall show below, for an
external regulating gene, the covariance, $\xi_{n,i}^e$ satisfies
\begin{equation} \label{corrext}
\xi_{n,i}^e \ge 0. 
\end{equation} 
The negatively self-regulating gene might have the covariance,
$\xi_{n,i}^s$, to be 
\begin{equation} \label{corrslf}
\xi_{n,i}^s \ge 0, \ \ {\rm or} \ \ \xi_{n,i}^s < 0. 
\end{equation}
Thus, while an externally regulated gene operates only on the
super-Poisson and Poisson regimes, the negative self-regulated gene
operates on both regimes plus the sub-Poissonian.

A closed form for Eqs. (\ref{corrext}) and (\ref{corrslf}), for the
binary gene, is written with the help of the exact solutions of the
Eqs. (\ref{equa1}) and (\ref{equa2})
\cite{Hornos05,innocentini07}. Before writing the closed forms for the
covariance for the externally regulated or self-interacting gene, we
redefine the model's parameters as:
\begin{eqnarray}\label{params} 
z_0 &=& \frac{\rho}{\rho+h_1}, \ \ N_1 = \frac{k}{\rho}, \ \ a =
\frac{f}{\rho}, \nonumber \\ 
b &=& \frac{f+h_2}{\rho}z_0 \,+\,
N_1z_0(1-z_0).
\end{eqnarray}
Particularly, one takes $z_0=1$ (or $h_1=0$) and it results
\begin{equation} \label{bext} 
b=\frac{f+h_2}{\rho}, 
\end{equation}
for the externally regulated gene and $0 \le a \le b$ when one compares the
Eqs. (\ref{params}) and (\ref{bext}). The negative self-regulating
gene has $h_2=0$ and
\begin{equation}\label{bself} 
b=[a + N_1(1-z_0)] z_0.
\end{equation} 
In the set of parameters $\{a,b,z_0\}$ one takes $N_1 =
\dfrac{b-az_0}{z_0(1-z_0)} $ and, due to positivity of $N_1$, we see
that, for a fixed $b$, $a$ lies in the interval $[0, b/z_0]$. The
choice for a fixed $b$ is because it is the invariant characterizing
the Lie symmetry of the binary model \cite{ramos07, ramos10}. Now we
present explicit forms for the covariances, which demonstration is
given at the supplementary material.

The covariance of the extenally regulated gene ($z_0=1$) is given by 
\begin{equation}
\xi^e_{n,i} = \frac{N_1^2}{b+1}\frac{a}{b}\frac{b-a}{b} \ge 0,
\end{equation}
since $a\le b$. 

For negatively self-regulating gene ($h_2=0$) the covariance has a
more complicated form, given in terms of the protein mean number
$\langle n \rangle = C N_1 (a z_0 / b) {\rm M} (a+1, b+1, N_1
z_0(1-z_0))$. Namely it is given as:
\begin{equation}
\xi^s_{n,i} = az_0\frac{N_1 - \langle n \rangle}{1-z_0} - \langle n
\rangle^2,
\end{equation}
where we used the Eq. (\ref{params}). The reader should keep in mind
that the average protein number is given as a function of the parameters
$(a,b,z_0)$ and $\langle n \rangle$ is used for shortness.

\begin{figure*}
\includegraphics[width = 0.23 \linewidth]{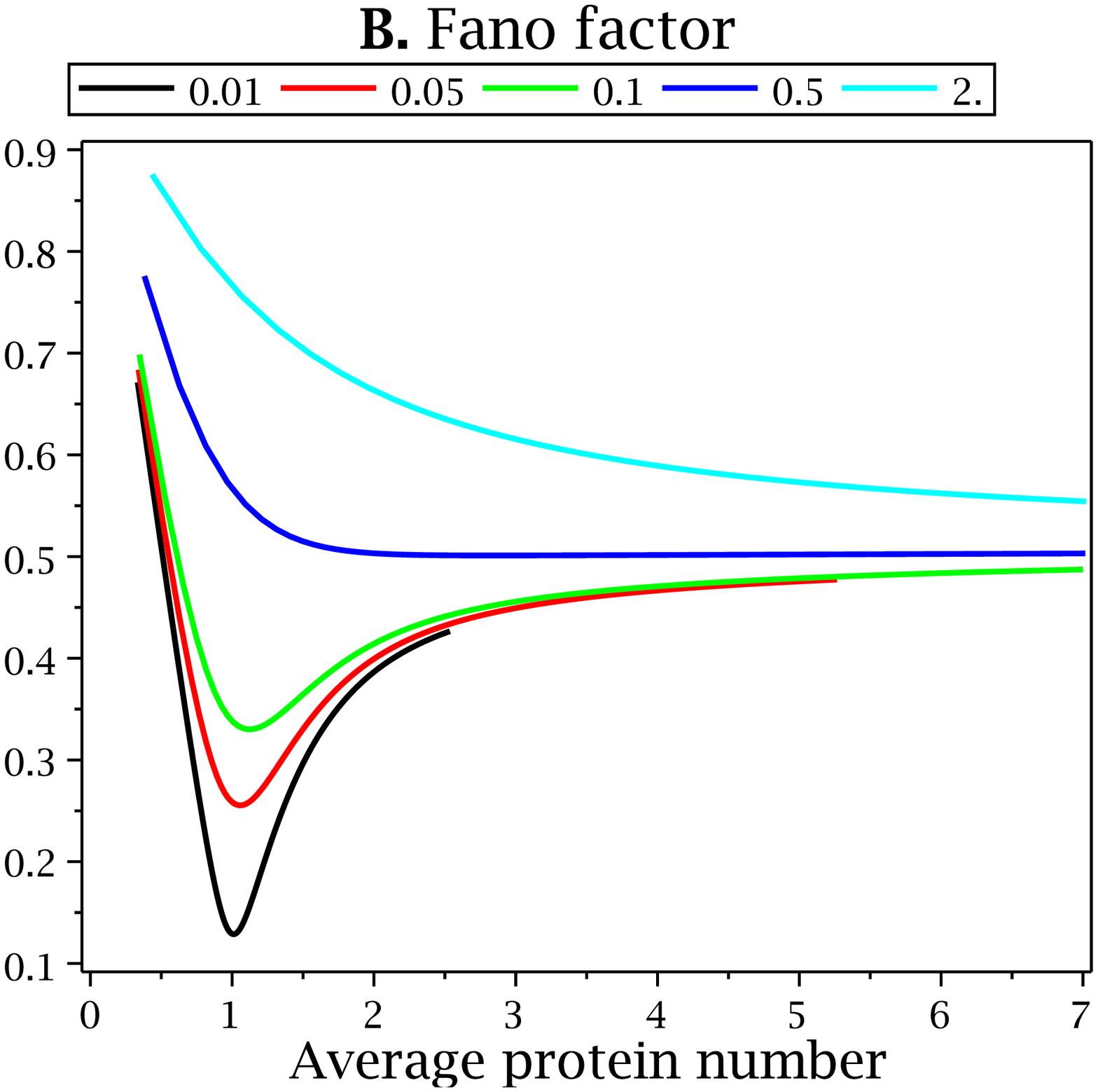} \ \ \ 
\includegraphics[width = 0.23 \linewidth]{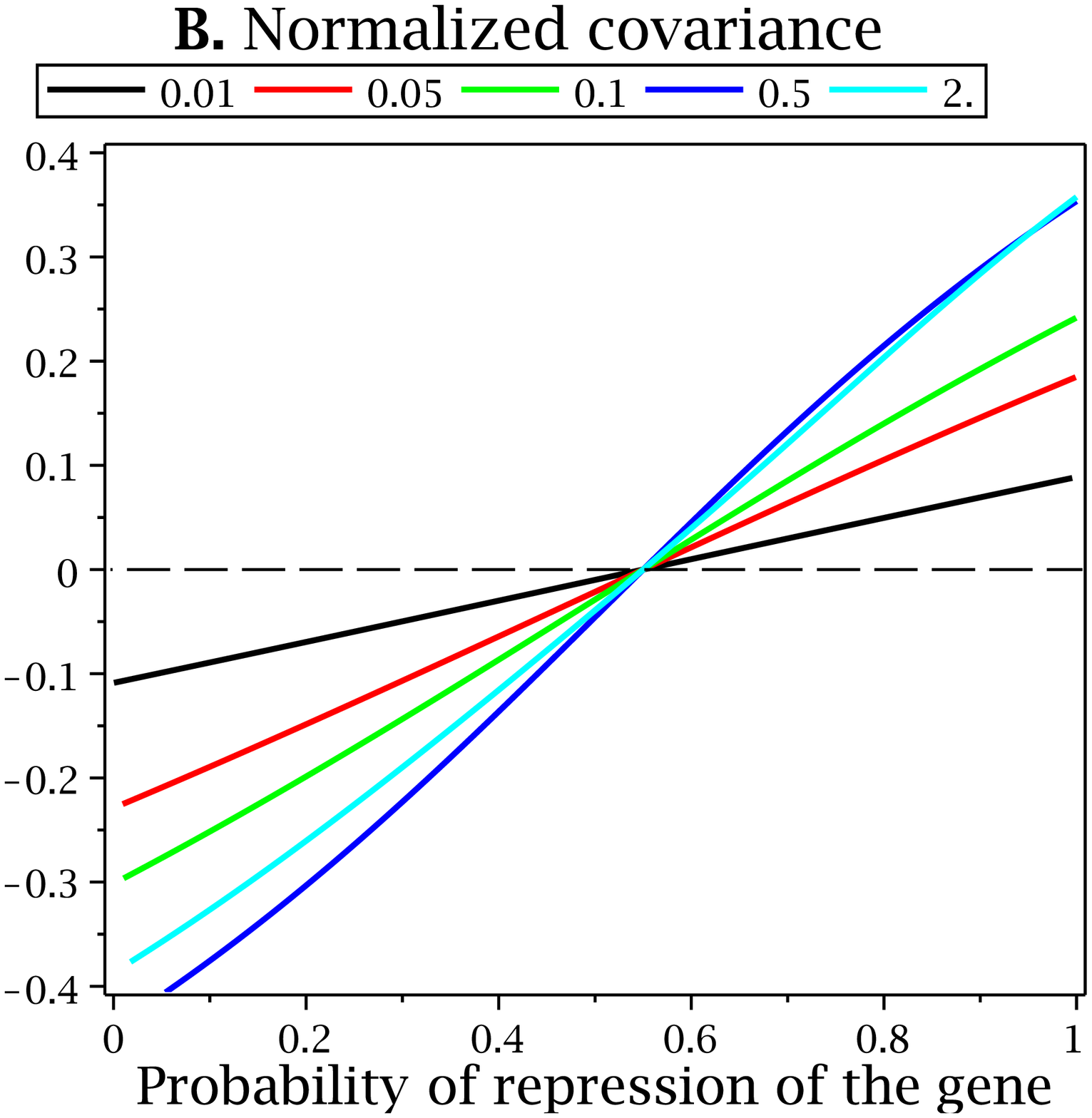} \ \ \ 
\includegraphics[width = 0.23 \linewidth]{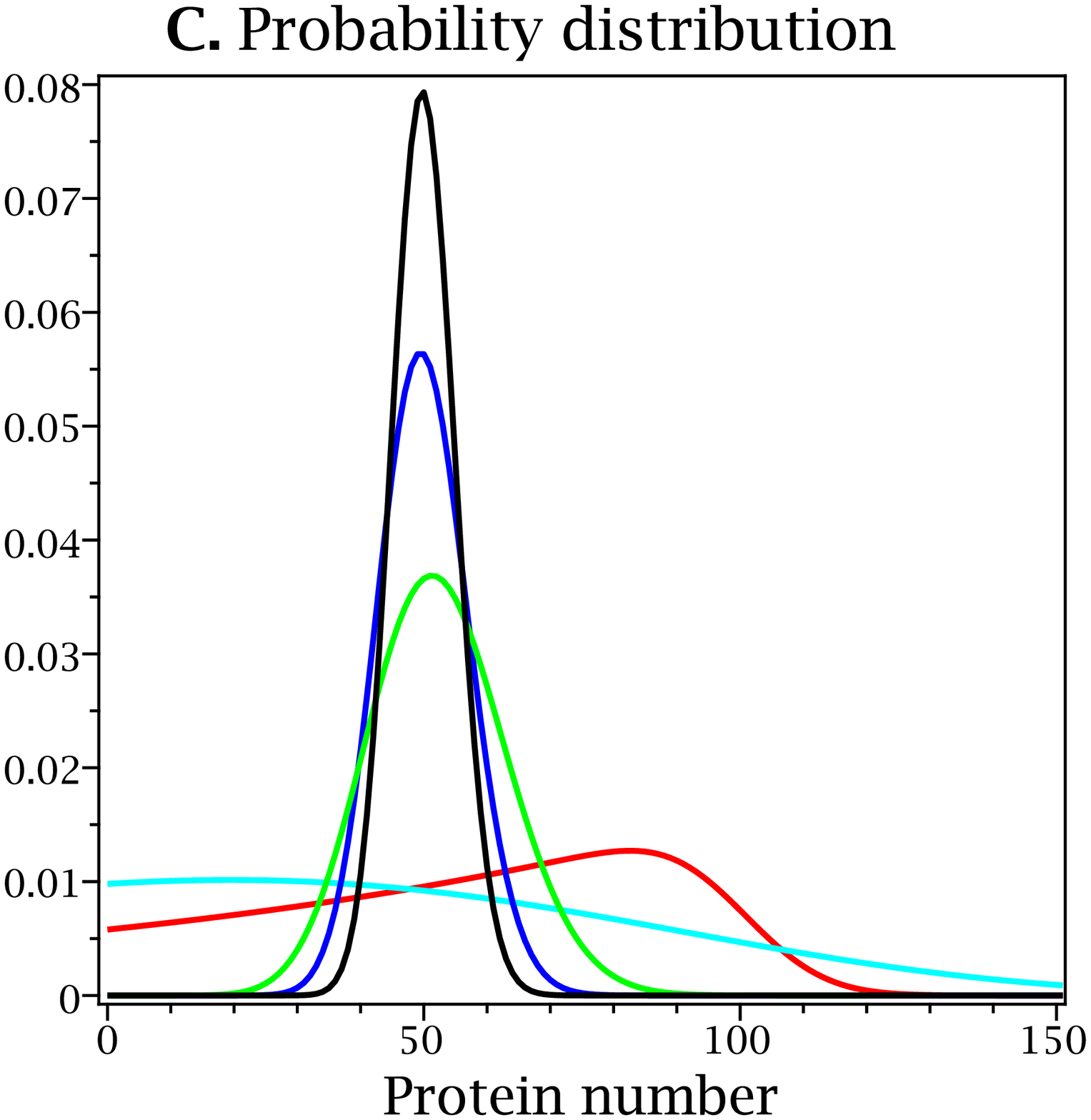}
\caption{{\bf A. Fano factor versus the average protein number.} Here
  we fixed $a=500$ with different colors standing for fixed values of
  $b$ as indicated on the legend. The Fano factor for the binary gene
  has 0.5 as an assymptotic limit, when the protein numbers tend to
  their maximum value, when $z_0 \rightarrow 0$. This limit
  corresponds to the condition of protein degradation very slow. The
  behaviour of $\mathcal{F}$ is dependent on the value of $b$ when
  $z_0 \rightarrow b/a$. It has a simple decay to 0.5 for $b>0.1$
  while it presents a pronounced minimum at $\langle n \rangle = 1$
  for the values of $b\le 0.1$. Surprisingly, the lowest value for the
  Fano factor occurs on the limit of low molecules number. {\bf
    B. Normalized covariance versus the probability to the gene to be
    active.} At that graphic we have plotted the Eq. (\ref{corr})
  versus $p_2$. We have fixed $z_0=0.45$ and, for each value of $b$,
  we varied $a$ from zero to $b/z_0$. The correlation is positive for
  $a<b$ and negative for $a>b$. For $a=b$ we have $p_2=1-z_0$ and the
  correlation is null. {\bf C. Probability distribution of the protein
    number} for the negatively self-regulating gene ($\phi_n = C
  \frac{(a)_n}{(b)_n}\frac{(N_1z_0)^n}{n!} {\rm M}(a+n, b+n,
  -Nz_0^2)$). The coupling between the ``on'' and ``off'' states
  sharpens the probability distribution as shown by the increase of
  the parameters $a$ and $z_0$. The constants ($a$, $b$,
  $z_0$) for the lines in black, blue, green, cyan and red, are,
  respectively, ($5\times10^3$, 1., $10^{-4}$), (50., 50., 0.5), (14.,
  70., 0.5), (1., 15., 0.95), (1., 2., 0.99).}
\label{fig1}
\end{figure*}

Fig. (\ref{fig1}) {\bf A} shows the Fano factor versus the average
protein number, for a fixed value of $a$, for the negative
self-interacting gene. Each line corresponds to a fixed value of $b$
and variation of $z_0$. As it have been demonstrated earlier
\cite{ramos07}, the sub-Fano regimes occur only when $a>b$. Since we
are exclusively interested on the sub-Fano regime we have investigated
only the condition $a > b$, that corresponds to the negative
covariance regime. Unexpectedly, there is a minimum value for the Fano
factor for the mean protein number equals to one. The Fano factor, for
the higher values of the average protein number, is smaller than the
Fano factor for the Poisson process by a factor 2.

Fig. (\ref{fig1}) {\bf B} shows the covariance, as calculated from the
ratio between the covariance by the variance on $n$ and the variance
on $N$, namely
\begin{equation} \label{corr}
\xi^s_{i,n}/\sigma/\sqrt{N_1^2p_1p_2},
\end{equation}
versus the probability for the gene to be active, $p_1$. The
covariance is positive for $p_1 < z_0$ (or $a<b$), when the gene has a
high probability to stay repressed. The covariance is zero for
$p_1=z_0$ (or $a=b$) and $\phi_n$ shall be a Poissonian
distribution. For the condition when $p_1>z_0$ (or $a>b$), the
covariance is negative and one obtains a sub-Fano regime, with located
probability distributions.

Fig. (\ref{fig1}) {\bf C} shows the effect of increasing the intensity
of coupling of the two gene states onto the probability distribution
$\phi_n$ of finding $n$ proteins inside the cell in the negatively
self-regulating gene. The spreader distributions, as shown by the
colors cyan and red, correspond to the condition of $a$ and $z_0$,
respectively, close to 0 and 1, that is the limit of low values for
the switching rate (or coupling) constants $f$ and $h_1$ at the
Eqs. (\ref{equa1}) and (\ref{equa2}). For intermediary values of the
coupling constants, the distribution is still super-Poisson and gets
thinner, as shown by the green colored line. A Poisson distribution is
represented by the blue line. The limit when $a>b$ and $z_0 \sim 0$,
for strong coupling, the probability gets highly located, as indicated
by the black line.

In the case of a sub-Poissonian process of protein synthesis, the Fano
factor is smaller than that of the uncoupled Poissonian
processes. Hence, the negative covariance induces canalization when
two stochastic processes are coupled. We suggest this as the
theoretical mechanism underlying the higher precision of the negative
self-regulating gene \cite{becskei00}: the possibility of regimes
where negative covariance between protein number and gene
synthesis rates (or, equivalently, gene states) exists.

Finally, the variance, $\sigma^2$, on the number of products of a gene
operating in multiple modes of synthesis follows directly from the
expression for the Fano factor as
\begin{equation} \label{variance}
\sigma^2 = \langle n \rangle + \xi_{n,i}.
\end{equation}
It is clear from this equation that the variance for the negatively
self-regulating binary gene is smaller than that of Poissonian
distribution, for a fixed average protein number, when the covariance
between the protein number and the gene state is negative. In other
words, the probability distribution on $n$ shall be highly located
around $\langle n \rangle$.

The existence of a negative covariance on a negatively self-regulating
gene is intuitively predicted from the analyzis of the
Eqs. (\ref{equa1}) and (\ref{equa2}). The coupling between these two
equations is given as a function of $n$. That is interpreted as
follows, the higher the number of proteins in the cytoplasm of the
cell, the higher the probability for the gene to switch to the
repressed state and, consequently, to have a lower value for the
protein synthesis rate, {\em i.e.}, an increase in $n$ might decrease
$N$. Despite not easy to demonstrate, it is tempting to conjecture the
existence of negative covariance regimes in negatively self-regulating
genes operating in more than two modes of expressio.

Biologically, the cell processes requiring higher precision would have
a biochemical machinery that implement the negative
covariance. Furthermore, one would expect the gene to switch multiple
times during a time interval without a significant change of the
protein number. Under these two assumptions it is expectable that the
variance on the protein number to be small.

In summary, in this manuscript we have shown that the higher precision
on the number of gene products by the stochastic gene under negative
self-regulation is due to the negative covariance between two random
variables: protein number and protein synthesis rate. Our results
suggest this as a general mechanism underlying the variance reduction
(or canalization) in the cell environment. Further research should
enlighten the biochemical implementation of negative covariance in
networks or cascades of biochemical reactions. Experimental
verification of our results would employ detection of both gene
activation and protein numbers and analysis of their covariance.

\end{document}